\documentclass[prd, twocolumn,    floatfix, superscriptaddress]{revtex4}
\usepackage{graphicx,subfigure,bm,color,psfrag,hyperref}
\usepackage{amsfonts}
\usepackage{lipsum}
\usepackage{mathtools}
\usepackage{verbatim}
\usepackage[normalem]{ulem}
\usepackage[dvipsnames]{xcolor}
\hypersetup{colorlinks,linkcolor={blue},citecolor={red},urlcolor={teal}}

\begin{document}

\title{Cosmological tensions in the era of precision cosmology: Insights from 
Tensions in Cosmology 2025}

\author{Eleonora Di Valentino}
\email{e.divalentino@sheffield.ac.uk}
\affiliation{School of Mathematical and Physical Sciences, University of Sheffield, Hounsfield Road, Sheffield S3 7RH, United Kingdom}

\author{Jackson Levi Said}
\email{jackson.said@um.edu.mt}
\affiliation{Institute of Space Sciences and Astronomy, University of Malta, 
Malta, MSD 2080}
\affiliation{Department of Physics, University of Malta, Malta}

\author{Emmanuel N. Saridakis}
\email{msaridak@noa.gr}
\affiliation{National Observatory of Athens, Lofos Nymfon, 11852 Athens, 
Greece}
\affiliation{CAS Key Laboratory for Researches in Galaxies and Cosmology, 
Department of Astronomy, University of Science and Technology of China, Hefei, 
Anhui 230026, P.R. China.}
\affiliation{Departamento de Matem\'{a}ticas, Universidad Cat\'{o}lica del 
Norte,  Avda. Angamos 0610, Casilla 1280 Antofagasta, Chile}

\begin{abstract}

The ``Tensions in Cosmology'' series of conferences has been established 
as one of the main venues where the cosmological community collectively 
assesses the cracks in the concordance model and explores possible theoretical 
and observational remedies. The 2025 edition, held once again in Corfu, Greece, 
came at a crucial time: the Hubble constant $H_0$ discrepancy has now exceeded 
$6\sigma$, and new high-precision data from DESI, JWST, ACT, and other facilities 
have made this tension more robust while opening new windows on the early and 
late Universe. The $S_8$ tension, though milder and survey-dependent, remains 
an important probe of late-time structure formation, while emerging anomalies 
involving dynamical dark energy and neutrino physics are gaining increasing 
attention as potential signs of physics beyond $\Lambda$CDM. Here we provide a 
report on the meeting and an update on the state of the tensions in 2025, 
highlighting progress since the pioneering 2022 event.

\end{abstract}

\maketitle
 
\section{Introduction}

The ``Tensions in Cosmology 2025'' Conference, co-organized by the Corfu Summer Institute and the National Observatory of Athens, gathered more than 130 participants from five continents, including leading experts in observations, theory, and data analysis. The meeting retained its now well-established blend of high-level plenary talks, focused contributed sessions, and lively discussions that extended well beyond the lecture hall. Five days of intense sessions were devoted to assessing the most persistent cosmological tensions~\cite{CosmoVerseNetwork:2025alb}, ranging from the $H_0$ and $S_8$ discrepancies to large-scale CMB anomalies, neutrino mass bounds, and beyond-$\Lambda$CDM scenarios~\cite{CosmoVerseNetwork:2025alb}. As in previous editions, the conference also attracted a large number of early-career researchers and students, confirming that the study of cosmological tensions remains a major driver for the next generation of cosmologists.

Compared to 2022, the field has entered a genuine ``precision tension'' era: new data releases such as DESI DR2, updated strong-lensing time-delay cosmography (TDCOSMO-2025), James Webb Space Telescope (JWST) Cepheid observations, and improved Tip of the Red Giant Branch (TRGB) calibrations, have made the $H_0$ discrepancy both more robust and more difficult to dismiss as a statistical fluctuation or systematic artifact. Similarly, the $S_8$ tension, far from fading, remains statistically significant when combining DES and HSC weak-lensing surveys, while KiDS-Legacy and eROSITA results suggest a milder or absent tension. This has motivated more sophisticated baryonic feedback modeling, improved descriptions of nonlinear structure growth, and advanced cosmic shear reconstruction techniques. Thus, the conference served as a snapshot of a field in which tensions are no longer peripheral curiosities but central, data-driven challenges to the concordance model.

Beyond the headline discrepancies, the 2025 edition reflected the 
broadening of the ``tensions'' program into a wider research agenda. 
Sessions were devoted to testing the foundations of $\Lambda$CDM, 
including BAO likelihood non-Gaussianities, Gaussian-process reconstructions of $H(z)$, 
and direct measurements of peculiar velocity fields. A growing synergy with 
gravitational-wave cosmology was also evident, with several talks demonstrating how 
standard sirens, pulsar-timing-array data, and stochastic gravitational-wave 
background constraints can provide independent tests of the expansion history 
and early-Universe physics. Overall, the conference offered not only a progress 
report on the status of key cosmological tensions, but also a vision for how the 
community intends to address them with the wealth of high-precision data 
anticipated in the second half of the decade.

\section{Progress on the Hubble Tension}

\begin{figure*}[t!]
    \centering
    \hspace{-2.5cm}
    \includegraphics[scale=0.8, angle=-90]{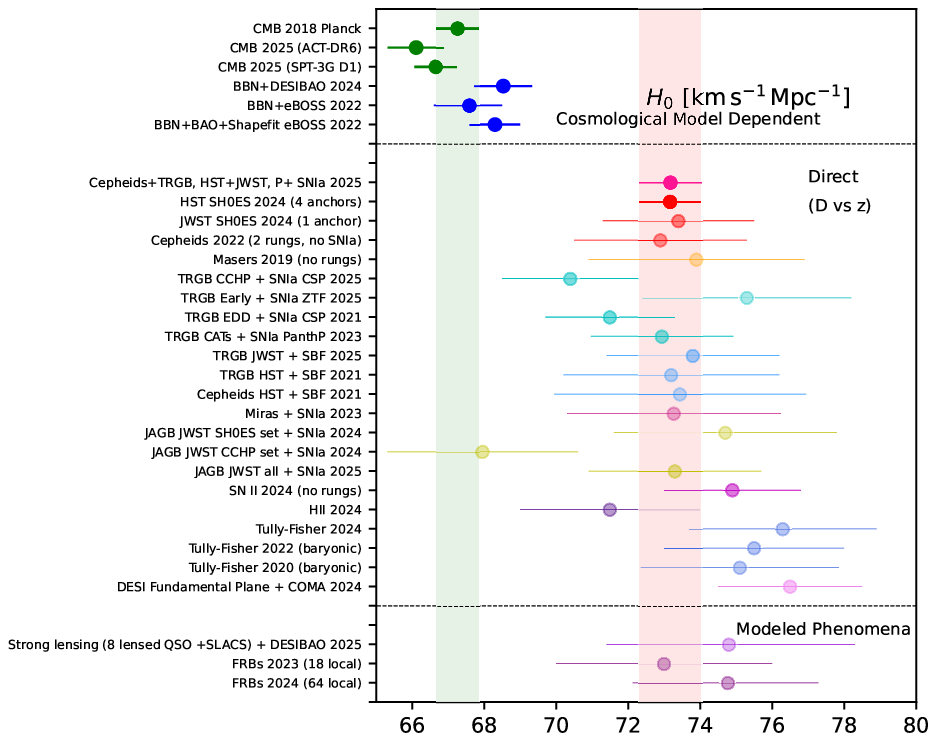} 
    \caption{{\it{Recent determinations of the Hubble constant $H_0$ from a 
variety of methods. Local distance-ladder approaches, including Cepheid- and 
TRGB-calibrated Type Ia supernovae, surface-brightness fluctuations, Type II 
supernovae, the Tully–Fisher relation, Mira variables, carbon stars, 
strong-lensing time-delay cosmography, fast radio bursts, the DESI fundamental 
plane with the Coma cluster, and maser distances, consistently favor $H_0 \simeq 
71$--$77~{\rm km\,s^{-1}\,Mpc^{-1}}$. In contrast, early-Universe inferences 
from the CMB and BAO within $\Lambda$CDM yield lower values, around $H_0 \simeq 
66$--$68~{\rm km\,s^{-1}\,Mpc^{-1}}$. Updated 
from~\cite{CosmoVerseNetwork:2025alb}.}}}
    \label{fig:H0}
\end{figure*}

The Hubble constant $H_0$ remained one of the main focal points of discussion and a 
major area of theoretical and observational innovation. Four years after the first 
Corfu conference, the tension between early- and late-Universe determinations 
has not subsided; rather, it has become more robust thanks to significant 
improvements in both local and high-redshift measurements (see Fig.~\ref{fig:H0}). 

On the local side, new distance-ladder results incorporating JWST Cepheid 
observations, Miras, and carbon-star calibrations have strengthened the SH0ES 
determination ($H_0 \simeq 73.17 \pm 0.86~{\rm km\,s^{-1}\,Mpc^{-1}}$), with 
additional robustness tests addressing crowding, metallicity, and photometric 
systematics. These efforts have reached a level of maturity where further 
reductions in statistical errors are likely to be incremental, underscoring the 
importance of cross-validation with completely independent approaches. A key 
highlight of the conference was the presentation of the latest TDCOSMO-2025 
blind analysis, which, using eight strongly lensed quasars and improved stellar 
kinematics from JWST, Keck, and VLT, yielded 
$H_0 = 72.1^{+4.0}_{-3.7}~{\rm km\,s^{-1}\,Mpc^{-1}}$. This result is fully 
consistent with the Cepheid-calibrated ladder and further strengthens the case 
for a high local $H_0$ value. Complementary probes such as the Maser Cosmology 
Project and surface-brightness-fluctuation distances also report concordant 
values, reinforcing the inference that the tension is not a mere artifact of a 
single method.

On the early-Universe side, DESI DR2 BAO measurements at $z \gtrsim 1$ remain in 
excellent agreement with the $\Lambda$CDM sound horizon inferred from Planck 
data, and the combined CMB constraints from Planck+SPT+ACT yield 
$H_0 = 67.24 \pm 0.35~{\rm km\,s^{-1}\,Mpc^{-1}}$. This makes it increasingly 
difficult for purely late-time new physics to reconcile the two regimes. This 
finding places significant pressure on model-builders, since any viable solution 
must either reduce the sound horizon at recombination (as in early dark energy 
or interacting neutrino scenarios) or involve a local departure from cosmic 
homogeneity (as in void models) without conflicting with other datasets. Indeed, 
new analyses of ``local void'' scenarios were presented, with bulk-flow 
reconstructions and number-count surveys lending partial support to the 
existence of a $\sim 300$ Mpc underdensity, though questions remain about the 
required depth and isotropy.

On the other hand, novel theoretical avenues were actively debated. 
Yukawa cosmology, which relates the graviton Compton wavelength to a 
redshift-dependent $H_0$, was proposed as a way to interpret the tension 
as a manifestation of quantum-gravitational limits on cosmological 
measurements. Another striking idea involved primordial black holes (PBHs) 
acting as natural drivers of an early dark energy phase: under certain mass 
ranges and abundances, PBH evaporation can temporarily increase the expansion 
rate before Big Bang Nucleosynthesis, thereby reducing the sound horizon and 
raising the inferred $H_0$ value.

Additionally, model-building efforts converged on minimal yet physically 
motivated modifications to the standard paradigm. One of the most discussed was 
the sign-switching cosmological constant scenario ($\Lambda_s$CDM), in which a 
late-time Anti-de Sitter to de Sitter transition occurs around $z \sim 1.7$. 
This framework, related to the concept of graduated dark energy (gDE), shows 
promising potential to simultaneously address both the $H_0$ and $S_8$ tensions 
while remaining consistent with CMB and BAO constraints. Several talks reported 
updated parameter constraints, showing that the transition redshift and width 
are now strongly bounded by DESI and Pantheon+ data, thereby providing a clear 
target for future surveys to confirm or rule out. Interacting dark energy–dark 
matter models were also revisited: weak but nonzero coupling strengths 
($\epsilon \sim 10^{-4}$) are mildly preferred by DESI low-redshift BAO data, 
suggesting that energy transfer from dark matter to dark energy might slightly 
enhance late-time expansion. Meanwhile, modified gravity frameworks such as 
$f(T)$ and $f(Q)$ cosmology have now reached the level of precision cosmology, 
with Bayesian model comparisons indicating that some of these scenarios can fit 
the combined dataset with a comparable, or in some cases slightly better, 
likelihood than $\Lambda$CDM, particularly when allowing for small spatial 
curvature or dynamical transitions.

Finally, an important message emerging from this discussion was that the 
``systematics hypothesis'', the idea that unknown errors would eventually 
erase the tension, is no longer the dominant view among practitioners. After 
more than a decade of scrutiny, with multiple independent probes all pointing 
in the same direction, the community is increasingly embracing the possibility 
that the Hubble tension is most likely a sign of new physics. The challenge now 
is to identify a minimal and testable extension of the concordance model that 
can accommodate the growing wealth of high-precision data without introducing 
new inconsistencies elsewhere.

\section{Status of the $S_8$ and Growth Tensions}

The $S_8$ tension, while somewhat less severe than the $H_0$ discrepancy, 
remained a major focus of the 2025 conference and is increasingly recognized as 
a complementary window into possible new physics. Weak-lensing experts presented 
the latest three-by-two point (3$\times$2pt) analyses combining KiDS-1000, DES 
Year~3, and HSC survey data, which consistently favor lower values of $S_8$ than 
those predicted by Planck+$\Lambda$CDM, with some combinations exceeding 
the $3\sigma$ level. These results have now reached a high degree of statistical 
robustness, with residual systematics such as shear calibration biases, 
photometric redshift errors, and intrinsic alignments carefully quantified, 
making it unlikely that they alone can explain the discrepancy.

A strong emphasis was placed on multiprobe consistency analyses, which 
cross-correlate Planck PR3/PR4 and ACT+WMAP CMB datasets with DESI BAO, 
galaxy clustering, and CMB lensing data. These combined analyses were shown to 
be powerful tools for diagnosing whether the source of the tension arises from 
astrophysical systematics, modeling choices (e.g., baryonic feedback), or 
genuinely new physics. A key development was the use of joint-likelihood 
approaches with non-Gaussian covariance modeling, which more accurately capture 
the information content of current surveys and help to reduce parameter 
degeneracies, such as between $\sigma_8$ and $\Omega_m$.

An important point of the conference was the presentation of novel techniques 
to directly reconstruct the matter power spectrum from cosmic shear data, 
extending from the linear into the nonlinear regime. These reconstructions allow 
a clean comparison between theory and observation, helping to disentangle 
baryonic feedback effects - whose impact can mimic suppressed growth - from new 
gravitational physics. In addition, several talks focused on hydrodynamical 
simulation suites and emulators that can marginalize over feedback uncertainties, 
as well as machine-learning-based emulators capable of spanning large 
cosmological parameter spaces at high precision. These efforts are crucial for 
the interpretation of upcoming Stage IV surveys (Euclid, LSST, Roman), which 
will deliver order-of-magnitude improvements in lensing statistics.

On the theory side, several contributions focused on exploring modifications to 
General Relativity that could reconcile weak-lensing results with CMB 
predictions. Teleparallel gravity models and other torsional modifications, as 
well as $f(Q)$ and scalar-tensor theories, were shown to predict suppressed 
growth rates at late times, bringing them into closer agreement with KiDS and 
DES measurements. Some of these models achieve this without significantly 
altering the CMB anisotropy spectrum, making them particularly attractive as 
potential explanations for the growth tension. Another avenue explored was the 
concept of dynamical dark energy with time-varying equation-of-state parameters 
($w_0, w_a$), which can alter the growth history and thus shift $S_8$. While 
many such models are now tightly constrained by DESI DR2 and Pantheon+ data, 
small departures from $w = -1$ remain allowed and could help alleviate the 
tension.

Furthermore, participants noted that the $S_8$ tension may not be a 
single phenomenon, but rather a manifestation of a more general mismatch in the 
late-time growth rate $f\sigma_8(z)$. Measurements of redshift-space distortions 
and peculiar velocity surveys (Cosmic-Flows) were discussed as independent probes 
of growth, with some showing a mild preference for lower growth than predicted 
by $\Lambda$CDM. Importantly, combining these probes with weak-lensing and cluster 
abundance measurements enables cross-validation of results and can clarify whether 
the tension persists across all tracers or is specific to lensing systematics.  

Overall, the consensus at the meeting was that the $S_8$ tension is now 
significant enough to warrant theoretical attention on par with the Hubble 
tension. Whether its resolution will come from improved modeling of small-scale 
baryonic physics, subtle extensions of $\Lambda$CDM, or entirely new 
gravitational degrees of freedom remains an open question. Nevertheless, the 
community is preparing for an huge amount of data from Euclid and LSST, which 
are 
expected either to confirm the tension at much higher significance or to reveal 
that it was driven by as-yet unaccounted-for systematics.

\section{New Directions and Emerging Anomalies}

Beyond the two major tensions, the conference featured a rich 
variety of sessions exploring what might be called the ``frontier anomalies'' of 
cosmology - effects that may represent hints of new physics or subtle 
challenges 
to our current assumptions. Large-scale CMB anomalies remained an active area of 
investigation, including the hemispherical power asymmetry, the lack of large-angle 
correlations, and alignments of low multipoles. Several talks examined whether these 
features could be explained by cosmic variance alone, or whether they instead point 
to a breakdown of statistical isotropy. New approaches highlighted the upcoming 
sensitivity of CMB-S4 and Simons Observatory data, as well as the potential of 
combining CMB maps with the cosmological gravitational-wave background (CGWB) to 
search for correlated signatures of early-Universe physics. The idea that CGWB data 
from next-generation interferometers could shed light on the origin of these anomalies 
generated considerable excitement, positioning gravitational-wave cosmology as a tool 
not only for inflationary science but also for probing late-time isotropy.

The role of astrophysical foregrounds also received renewed attention, with 
presentations on the possible contribution of massive elliptical galaxies to the 
observed CMB photon energy density. If confirmed, such a contribution could 
necessitate a partial reinterpretation of the CMB monopole and would have 
profound implications for the inferred cosmological parameters. Related works 
explored the possibility that dust emission and early star formation could 
imprint subtle spectral distortions, potentially biasing cosmological parameter 
inference if not properly accounted for.  

Neutrino physics emerged as another focal point, with several talks emphasizing 
the tension between cosmologically inferred upper bounds on the sum of neutrino 
masses and the lower limits implied by oscillation experiments. Some speakers 
argued that this constitutes an independent ``neutrino mass tension,'' which may 
hint at non-standard neutrino interactions, sterile species, or a need for 
beyond-$\Lambda$CDM physics such as evolving dark energy. Models were presented 
that attempt to simultaneously address the $H_0$, $S_8$, and neutrino mass 
anomalies, with varying degrees of success, underscoring the challenge of 
achieving a fully coherent picture.

Another methodological theme was the importance of rigorous and innovative 
statistical tools. Talks on non-Gaussian likelihood treatments for BAO showed 
that approximations used in past analyses could introduce mild biases in 
parameter inference in the era of percent-level precision. Gaussian-process 
reconstructions of the expansion history were presented as model-independent 
approaches to search for hints of new physics, such as transitions in $w(z)$ or 
variations of fundamental constants. Particularly notable were new 
forward-modeling frameworks such as GalSBI and SHAM-OT, which generate highly 
realistic mock galaxy catalogs by matching luminosity functions, morphologies, 
and spatial distributions with unprecedented fidelity. These tools are expected 
to play a crucial role in Stage IV surveys, ensuring that photometric redshift 
calibrations and selection effects do not artificially create or suppress 
cosmological tensions.

Finally, an important feature of the conference was the rapidly maturing synergy 
between cosmology and gravitational-wave astronomy. Multiple contributions 
explored how binary black hole and neutron star merger environments could serve 
as probes of dark matter halos, testing for dynamical friction and accretion 
signatures. The stochastic gravitational-wave background (SGWB), measured by 
pulsar-timing arrays such as NANOGrav, was extensively discussed as a potential 
window on phase transitions, cosmic strings, and inflationary reheating 
scenarios. Presentations quantified how future pulsar-timing array (PTA) 
sensitivity extending to $\mu$Hz frequencies could enable a $>3\sigma$ detection 
of extra radiation energy density ($\Delta N_{\rm eff}$), providing an entirely 
new perspective on early-Universe physics. Taken together, these developments 
suggest that the next decade will not only refine our understanding of existing 
tensions but may also uncover qualitatively new phenomena that could point the 
way toward the next cosmological paradigm.

\section{Conclusions}

Three years after the first Corfu meeting, the cosmological tensions have not 
dissipated; on the contrary, they have crystallized into one of the defining 
puzzles of modern physics. The $H_0$ tension now stands at over $6\sigma$, 
confirmed by a suite of independent techniques including Cepheid- and 
TRGB-calibrated distance ladders, strong-lensing time-delay cosmography, and 
maser measurements. Likewise, the $S_8$ and growth-rate tensions remain 
statistically significant, despite increasingly sophisticated treatments of shear 
calibration, photo-$z$ biases, and baryonic feedback modeling, or partial data 
releases that suggest milder discrepancies. These persistent disagreements have 
shifted the conversation from one of skepticism to one of cautious recognition 
that we may be probing cracks in the $\Lambda$CDM paradigm itself. Theoretical 
creativity is flourishing, with ideas ranging from interacting dark sectors and 
early dark energy to quantum-gravity-inspired modifications of the Friedmann 
equations. However, no single proposal has yet achieved broad consensus or fully 
resolved all tensions and anomalies simultaneously.

Another important development since 2022 is the community’s growing emphasis on 
global and multiprobe consistency tests. It is no longer sufficient to focus on 
a single dataset or probe, as progress now comes from the joint analysis of CMB, 
BAO, SNe~Ia, weak lensing, RSDs, and strong-lensing datasets within a common 
statistical framework. This integrated approach has already begun to reveal 
subtle tensions that would remain hidden in isolation, while also clarifying 
where residual systematics may still play a role. In parallel, new observational 
frontiers are opening rapidly: gravitational-wave cosmology promises independent 
distance measurements through standard sirens, while high-redshift quasar samples 
and forward-modeled galaxy surveys are extending the reach of the Hubble diagram 
and growth measurements deeper into the cosmic past. Taken together, these 
developments mark a maturation of the field into a genuinely data-driven, 
cross-validated science of cosmic consistency.

As in 2022, the atmosphere at the Corfu conference combined intellectual 
intensity with optimism: participants debated vigorously but constructively, and 
the message that emerged was clear - the cosmological community is not only 
ready 
but eager to face the possibility of a paradigm shift if the data demand it. The 
next decade, with Euclid, Roman, LSST, SO, LiteBIRD, and third-generation 
gravitational-wave detectors, will likely be decisive. Either the tensions will 
fade under the weight of next-generation data, restoring confidence in 
$\Lambda$CDM, or - what now seems the more probable outcome - the tensions will 
be 
confirmed at even higher significance, compelling a reformulation of our 
standard cosmological model. In either case, the ``Tensions in Cosmology'' 
conference series has established itself as a catalyst for progress, providing a 
forum where observers, theorists, and statisticians converge to chart the path 
forward in what may prove to be a new golden era of cosmology.


\end{document}